\documentclass[conference]{IEEEtran}
\IEEEoverridecommandlockouts

\usepackage{cite}
\usepackage{amsmath,amssymb,amsfonts}
\usepackage{algorithmic}
\usepackage{graphicx}
\usepackage{textcomp}
\usepackage{xcolor}
\def\BibTeX{{\rm B\kern-.05em{\sc i\kern-.025em b}\kern-.08em
    T\kern-.1667em\lower.7ex\hbox{E}\kern-.125emX}}

\usepackage[utf8]{inputenc}

\begin{document}

\title{Mamba-SEUNet: Mamba UNet for Monaural Speech Enhancement\vspace{-1mm}}

\author{
  \IEEEauthorblockN{Junyu Wang$^{1,*}$, Zizhen Lin$^{2,*}$, Tianrui Wang$^1$, Meng Ge$^1$, Longbiao Wang$^{1,3,\dagger}$, Jianwu Dang$^4$}
  \vspace{-0.1mm} 
  \IEEEauthorblockA{
    $^1$Laboratory of Cognitive Computing and Application, College of Intelligence and Computing, Tianjin University, \\Tianjin, China,
    $^2$School of Electronic Information, Sichuan University, Sichuan, China, $^3$Huiyan Technology (Tianjin) \\Co., Ltd, Tianjin, China, 
    $^4$Shenzhen Institute of Advanced Technology, Chinese Academy of Sciences, China\\
    \vspace{-13.5mm} 
  }
  \thanks{$^{*}$Equal contribution to this work. $^{\dagger}$Corresponding author.}
  \thanks{This work was supported by the National Natural Science Foundation
of China under Grant 62176182 and U23B2053.}
}


\maketitle

\begin{abstract}
In recent speech enhancement (SE) research, transformer and its variants have emerged as the predominant methodologies. However, the quadratic complexity of the self-attention mechanism imposes certain limitations on practical deployment. Mamba, as a novel state-space model (SSM), has gained widespread application in natural language processing and computer vision due to its strong capabilities in modeling long sequences and relatively low computational complexity. In this work, we introduce Mamba-SEUNet, an innovative architecture that integrates Mamba with U-Net for SE tasks. By leveraging bidirectional Mamba to model forward and backward dependencies of speech signals at different resolutions, and incorporating skip connections to capture multi-scale information, our approach achieves state-of-the-art (SOTA) performance. Experimental results on the VCTK+DEMAND dataset indicate that Mamba-SEUNet attains a PESQ score of 3.59, while maintaining low computational complexity. When combined with the Perceptual Contrast Stretching technique, Mamba-SEUNet further improves the PESQ score to 3.73.
\end{abstract}

\begin{IEEEkeywords}
speech enhancement, state space model, deep learning, mamba, u-net.
\end{IEEEkeywords}

\section{Introduction}
Speech enhancement (SE) tasks aim to improve speech clarity by suppressing background noise, reverberation, and other acoustic interferences, thereby optimizing user experience and communication efficacy. In recent years, with the rapid development of deep learning, a variety of representative neural networks have emerged, especially those based on convolutional neural networks (CNN) \cite{convtasnet, DDAEC, basedconv, CCAConv}, transformers \cite{DPT-FSNet, DBAIAT, DPCFCS}, and U-Net architectures \cite{S4NDU-Net, MANNER, MUSE}. Generally, depending on the processing method of the input signal, it can be broadly categorized into time-domain and time-frequency (T-F) domain approaches.

Time-domain studies enhance noisy speech signals directly in the time domain, leveraging encoder-decoder frameworks to map noise waveforms to clean ones \cite{waveformdomain1, ARFDCN}. While this approach retains all inherent information features of the time-domain signal, it often results in relatively coarse enhancements. In contrast, T-F domain SE methods demonstrate better capabilities in discerning the intricate structural details between speech and noise \cite{HGCN, DNS-RUI, 23interspeech}.

The current mainstream SE models primarily employ two-stage (TS) architectures based on Transformer or its variant, Conformer \cite{CMGAN, MPSENET}. By learning input features in both time-domain and frequency-domain representations, these models achieve outstanding SE performance. However, TS methods typically perform dimensionality reduction only along the frequency dimension, which makes it difficult to capture the characteristics of the input information at different resolutions, thereby constraining their performance ceiling to some extent. Moreover, the quadratic complexity of the self-attention mechanism with respect to sequence length poses significant challenges for deploying these models in scenarios with limited computational resources. Recent research \cite{MUSE} introduced MUSE, a more efficient approach combining sub-quadratic complexity Taylor multi-head self-attention (T-MSA) \cite{taylor} with U-Net framework \cite{U-Net}. By applying Taylor-Transformer to capture multi-scale information at different resolutions, MUSE achieves competitive performance with just 16 input channels and low computational complexity. However, the SE performance of T-MSA still falls short when compared to the original multi-head self-attention (MHSA) mechanism.

In parallel, developments in state-space models (SSM) \cite{S4NDU-Net, SSM} present a promising alternative with linear complexity and high efficiency in handling long-sequence inputs. Mamba \cite{mamba}, as a novel structured SSM (S4), introduces a selective processing mechanism for input information and an efficient hardware-aware algorithm, achieving performance comparable to or exceeding Transformer-based methods across domains such as natural language, image, and audio \cite{spmamba, mambaunet, mambavision}. Particularly, a recent work \cite{semamba} demonstrated improved performance with reduced FLOPs by simply replacing the conformer in MP-SENet with Mamba, further validating the effectiveness of Mamba in speech processing tasks.

\begin{figure*}[t]
  \centering
  \vspace{-0.4cm}
  \includegraphics[width=0.84\linewidth]{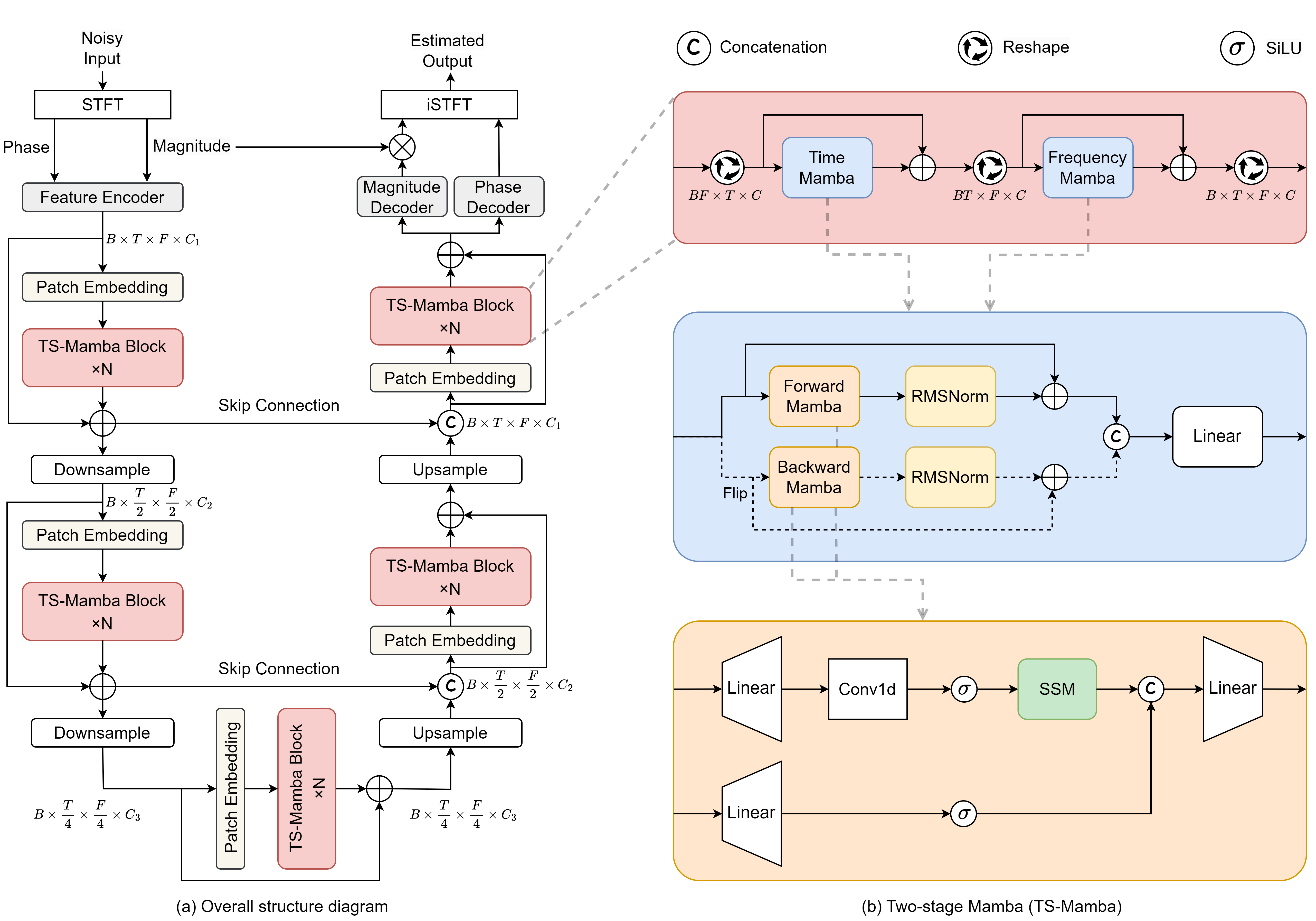}
  \caption{The architecture of the proposed Mamba-SEUNet.}
  \label{figure1}
  \vspace{-0.4cm}
\end{figure*}

Inspired by these studies, we propose Mamba-SEUNet, a model that integrates Mamba with U-Net. By leveraging TS-Mamba within the U-Net framework to learn coarse-grained and fine-grained information at different resolutions and performing multi-scale feature fusion, this network exhibits enhanced long-range dependency modeling and detail restoration capabilities. Extensive experiments on the VCTK+DEMAND dataset \cite{vctk+demand} demonstrate that Mamba-SEUNet achieves state-of-the-art (SOTA) performance, with a notable reduction in computational complexity measured in FLOPs.

\section{Mamba: Selective State Space Model}

As an extension of S4, Mamba introduces a selection mechanism that enables model parameters to be dynamically adjusted according to the inputs. This mechanism allows for the selective mapping of the input \(x(t)\) through a high-dimensional hidden state \(h(t)\) to the output \(y(t)\), which can be expressed as follows:
\begin{equation}
\begin{aligned}
h'(t) &= Ah(t) + Bx(t) \\
y(t) &= Ch(t) \\
\end{aligned}
\end{equation}
To apply to discrete speech signals, it is necessary to discretize \(A\) and \(B\) in the Eq.1. Specifically, given a time-scale parameter $\Delta$, the discrete parameter can be approximated using a zero-order hold as follows:
\begin{equation}
\begin{aligned}
\overline{A} = e^{(\Delta A)}, \thickspace \overline{B} = (\Delta A)^{-1}(e^{(\Delta A)}-I) \cdot (\Delta B) \\
\end{aligned}
\end{equation}
Thus Eq.1 can be rewritten as:
\begin{equation}
\begin{aligned}
h(t) &= \overline{A}h(t-1) + \overline{B}x(t) \\
y(t) &= \overline{C}h(t) \\
\end{aligned}
\end{equation}
Further expanding the computation in Eq.3 along the sequence, it can be seen that the output
\(y\) is computed from the input \(x\) through a global convolution with kernel  \(\overline{K}\):
\begin{equation}
\begin{aligned}
\overline{K} &= (C\overline{B},\overline{AB},...,C\overline{A}^{L-1}\overline{B}) \\
y &= x \ast \overline{K} \\
\end{aligned}
\end{equation}
where \(L\) is the size of the input sequence.

Additionally, a significant contribution of Mamba is its hardware-aware algorithm,  which enhances the efficiency of model execution on modern hardware. The core idea is to leverage the memory hierarchy of modern hardware, such as GPUs, to minimize I/O access between different memory levels. By executing discrete and recursive operations within the faster SRAM and returning final results to the slower HBM, the algorithm significantly reduces computations associated with \((B, L, D, N)\), where \(B\), \(L\), \(D\), and \(N\) represent the batch size, sequence length, number of channels, and state dimension respectively, thereby enhancing computational speed.

\section{Methodology}
\label{Mamba}

\subsection{Architecture Overview}

The overall structure of the proposed Mamba-SEUNet is illustrated in Fig. 1(a). Given the noisy speech input \(y\), we first apply the Short-Time Fourier Transform (STFT) to obtain its magnitude spectrum \(Y_m \in \mathbb{R}^{T \times F}\) and phase spectrum \(Y_p \in \mathbb{R}^{T \times F}\). These are then combined and transformed into an intermediate feature space \(Y \in \mathbb{R}^{T \times F \times C_1}\) via a feature encoder. Subsequently, these patch features are processed through a series of TS-Mamba blocks with skip connections, patch embedding layers, and downsampling and upsampling operations for hierarchical processing at different resolutions. The patch embedding layers, come from \cite{taylor}, leverage a combination of depthwise separable and deformable convolutions, enabling the capture of fine-grained acoustic details and enriching the subsequent TS-Mamba blocks with enhanced feature representations. The TS-Mamba blocks focus on learning these representations and capturing time and frequency dependencies. Finally, the enhanced magnitude and phase spectra are decoded and reconstructed into the enhanced speech signal using the inverse STFT (ISTFT).

\subsection{TS-Mamba Block}

We employ time and frequency Mamba blocks sequentially to learn the comprehensive feature representations, as illustrated in Fig. 1(b). To effectively capture both global and local information, each Mamba block incorporates the bidirectional SSM formulation proposed in \cite{BidirectionalSSM}, allowing the model to integrate both past and future information. This is similar to the forward and backward scanning mechanisms utilized in BLSTM \cite{BLSTM}. Specifically, for a given input \(x\), it is processed in parallel through both the forward Mamba and backward Mamba. Post-processing with RMSNorm \cite{RMSnorm} is applied to both the forward and backward outputs, after which they are concatenated and passed through a linear layer to obtain the output \(x'\):
\begin{equation}
\begin{aligned}
x_f &= RMS(FMamba(x)) + x \\
x_b &= RMS(BMamba(Flip(x))) + x \\
x' &= Linear(Concat(x_f, x_b)) \\
\end{aligned}
\end{equation}
where \(x_f\), \(x_b\), \(RMS\), \(FMamba\), \(BMamba\), \(Flip\), and \(Concat\) denote the forward output, backward output, RMS normalization, forward Mamba, backward Mamba, flip, and concatenation operations respectively.

The forward and backward Mamba blocks share an identical structure. Specifically, for an input sequence \( x_{in} \in \mathbb{R}^{L \times C} \), we first apply a linear layer to map it to \( x'_{in} \in \mathbb{R}^{L \times 2C} \). This is followed by a convolution operation with a kernel size of 4, coupled with a Sigmoid Linear Unit (SiLU) activation function. The output on one side \( x_1 \in \mathbb{R}^{L \times 2C} \) is then obtained using the SSM described in Eq.4. To compensate for any information loss due to the sequential constraints of the SSM, we add a symmetric gated branch without convolution and SSM, consisting of an additional linear layer and SiLU activation. Finally, the outputs from both branches are concatenated and projected through the last linear layer to obtain the output \( x_{out} \in \mathbb{R}^{L \times C} \), as expressed in the following equation:
\begin{equation}
\begin{aligned}
&x_1 = SSM(\delta(Conv(Linear(x_{in})))) \\
&x_2 = \delta(Linear(x_{in}))\\
&x_{out} = Linear(Concat(x_1 , x_2))\\
\end{aligned}
\end{equation}
where \(Conv\) denotes the 1-D convolution operation, $\delta$ is the SiLU activation function, and \(Concat\) represents the concatenation operation.

\subsection{Encoder and Decoder}

The encoder-decoder framework of Mamba-SEUNet is inspired by the methodology employed in MP-SENet \cite{MPSENET}. Its feature encoder consists of two convolutional layers and a Dilated DenseNet \cite{DenseNet}. The first convolutional layer increases the input channels to \(C_1\), producing an intermediate feature map, while the second halves the frequency dimension to optimize computational efficiency. In this study, the Dilated DenseNet is designed with a depth of 4 and a dilation factor of 2 to capture fundamental spectral features of speech. Both the magnitude and phase decoders incorporate the Dilated DenseNet structure from the encoder, followed by a 2-D transposed convolution and a \(1 \times 1\) convolution. The primary distinction between the two is the activation function: the magnitude decoder utilizes a learnable sigmoid function (L-Sigmoid) \cite{metricgan+}, whereas the phase decoder employs a two-argument arctangent function (Arctan2) \cite{MPSENET}.

\begin{table}[t]
  \caption{Hyperparameters for the Mamba-SEUNet XS, S, M, and L models}
  \label{tab1}
  \centering
  \vspace{-0.1cm}
  \tabcolsep=0.15cm  
  \begin{tabular*}{\hsize}{@{\hspace{0.3cm}}@{\extracolsep{\fill}}lccc@{\hspace{0.3cm}}}  
    \hline
    \textbf{Model} & \textbf{Dimension \(C_1\)} & \textbf{Blocks \(N\)} & \textbf{\# Params} \\
    \hline
    Mamba-SEUNet (XS)  & 16 & 2 & 0.99M  \\
    Mamba-SEUNet (S)  & 16  & 4  & 1.88M    \\
    Mamba-SEUNet (M)  & 24  & 4  & 3.78M    \\
    Mamba-SEUNet (L)   & 32  & 4  & 6.28M   \\
    \hline
  \end{tabular*}
  \vspace{-0.4cm}
\end{table}

\begin{table*}[ht]
\begin{center}
  \caption{Performance comparison on VCTK+DEMAND dataset. "-" denotes the data is not provided in the original paper.}
  \setlength{\tabcolsep}{4mm}
  \label{tab2}
  \centering
  \vspace{-0.1cm}
  \begin{tabular}{lccccccc}
    \hline
    \textbf{Model}   & \textbf{\# Params}  & \textbf{FLOPs}  & \textbf{WB-PESQ}  & \textbf{STOI}  &\textbf{CSIG}   & \textbf{CBAK}   & \textbf{COVL}  \\
    \hline
    Noisy  & - & - & 1.97 & 0.91 & 3.35 & 2.44 & 2.63                   \\
    \hline
    MetricGAN+ \cite{metricgan+} & - & - & 3.15 & - & 4.14 & 3.16 & 3.64                \\
    TSTNN \cite{TSTNN} & 0.92M & - & 2.96 & 0.95 & 4.33 & 3.53 & 3.67  \\
    CMGAN \cite{CMGAN}  & 1.83M & 63.15G & 3.41 & \textbf{0.96} & 4.63 & 3.94 & 4.12         \\
    DPCFCS-Net \cite{DPCFCS}  & 2.86M & 130.65G & 3.42 & \textbf{0.96} & 4.71 & 3.88 & 4.15  \\
    MP-SENet \cite{MPSENET} & 2.05M & 74.29G & 3.50 & \textbf{0.96} & 4.73 & 3.95 & 4.22          \\
    MUSE \cite{MUSE}  & 0.51M & 9.43G & 3.37 & 0.95 & 4.63 & 3.80 & 4.10            \\
    S4ND U-Net \cite{S4NDU-Net} & 0.75M & - & 3.15 & - & 4.52 & 3.62 & 3.85            \\
    SEMamba \cite{semamba} & 2.25M & 65.46G & 3.52 & \textbf{0.96} & 4.75 & 3.98 & 4.26            \\
    \hline
    Mamba-SEUNet (XS) & 0.99M & 4.16G & 3.48 & \textbf{0.96} & 4.75 & 3.95 & 4.23      \\
    Mamba-SEUNet (S) & 1.88M & 4.62G & 3.54 & \textbf{0.96} & 4.77 & 3.98 & 4.28      \\
    Mamba-SEUNet (M) & 3.78M & 10.28G & 3.57 & \textbf{0.96} & 4.79 & 4.00 & 4.30 \\
    Mamba-SEUNet (L) & 6.28M & 18.17G & \textbf{3.59} & \textbf{0.96} & \textbf{4.80} & \textbf{4.02} & \textbf{4.32}                   \\
    \hline
  \end{tabular}
  \vspace{-0.5cm}
\end{center}
\end{table*}

\section{Experiments}

\subsection{Datasets}

We employ the extensively adopted VCTK+DEMAND dataset \cite{vctk+demand} to evaluate the effectiveness of the proposed method. The clean speech data comes from the VoiceBank corpus \cite{voicebank}, while the mixed noise is sourced from the DEMAND dataset \cite{Demand}. The generated mixed speech dataset includes 12,396 utterances from 30 speakers, with 28 speakers used for training and 2 for testing. The training and testing sets include 10 types of noise with signal-to-noise ratios (SNRs) ranging from 0 to 15 dB and 5 types of noise with SNRs from 2.5 to 17.5 dB. All samples are downsampled to 16 kHz.

\subsection{Experimental setup}

During training, the speech data is uniformly segmented into 30600 points. The FFT length is set to 510, with a window length of 510 and a hop size of 120. The channel numbers \(C_2\) and \(C_3\) are set to \(\frac{C_1}{2}\) and \(\frac{C_1}{3}\), respestively. All models are trained for 200 epochs using the AdamW optimization \cite{adamw}, with an initial learning rate of 0.0005, decaying by a factor of 0.99 per epoch. Table \ref{tab1} provides the hyperparameters for the four models. 

\subsection{Evaluation metrics}

The efficacy of the proposed method is evaluated using various metrics. (1) Wide-band PESQ metric evaluates speech quality on a score scale from -0.5 to 4.5 \cite{pesq}. (2) Speech intelligibility is measured by STOI \cite{stoi}, with a score range from 0 to 1. (3) Mean opinion score (MOS) ratings include CSIG for signal distortion, CBAK for background noise interference, and COVL for overall speech quality, all rated from 1 to 5. (4) FLOPs are calculated based on processing a 2-second, 16kHz audio sample on a GPU.

\section{Results and analysis}

\subsection{Performance Comparison with Baselines}

Table \ref{tab2} presents a performance comparison between the proposed Mamba-SEUNet and several baseline models on the VCTK+DEMAND dataset. This comparison includes classical models such as MetricGAN+ and TSTNN, as well as state-of-the-art (SOTA) models like MP-SENet and SEMamba. The results indicate that Mamba-SEUNet (XS) achieves performance comparable to MP-SENet with just 0.99M parameters and 4.16G FLOPs. As the number of TS-Mamba blocks \(N\) increases to 4, Mamba-SEUNet (S) surpasses all existing models with 1.88M parameters and 4.62G FLOPs. By increasing the number of channels from 16 to 24, an improvement across all evaluation metrics is observed. Expanding the output dimension of the feature encoder to 32 further enhances the PESQ score of Mamba-SEUNet (L) to 3.59.
 
\begin{table}[h]
  \vspace{-0.3cm}
  \caption{Performance comparison between SEMamba and Mamba-SEUNet with PCS}
  \label{tab3}
  \centering
  \vspace{-0.1cm}
  \tabcolsep=0.11cm  
  \begin{tabular*}{\hsize}{@{\hspace{0.15cm}}@{\extracolsep{\fill}}lccccc@{\hspace{0.3cm}}}
    \hline
    \textbf{Model} & \textbf{WB-PESQ} & \textbf{STOI} & \textbf{CSIG} & \textbf{CBAK} & \textbf{COVL} \\
    \hline
    SEMamba + PCS \cite{semamba}  & 3.69 & \textbf{0.96} & 4.79 & 3.63 & 4.37 \\
    Mamba-SEUNet (S) + PCS & 3.70 & \textbf{0.96} & 4.79 & 3.64 & 4.37   \\
    Mamba-SEUNet (L) + PCS & \textbf{3.73} & \textbf{0.96} & \textbf{4.82} & \textbf{3.67} & \textbf{4.40} \\
    \hline
  \end{tabular*}
  \vspace{-0.2cm}
\end{table}

To provide a more comprehensive evaluation of model performance, we also report results using the Perceptual Contrast Stretching (PCS) technique \cite{pcs}, as shown in Table \ref{tab3}. PCS, a data preprocessing method, aims to enhance the perceptual quality of speech signals at the expense of background noise estimation. The results demonstrate that Mamba-SEUNet (S), with a similar number of parameters, achieves comparable performance with significantly lower FLOPs compared to SEMamba. By further increasing the number of channels, Mamba-SEUNet (L) improves the PESQ score to 3.73.

\begin{table}[h]
  \vspace{-0.3cm}
  \caption{Test results for different number of blocks \(N\) in Mamba-SEUNet (M).}
  \label{tab4}
  \centering
  \vspace{-0.1cm}
  \tabcolsep=0.11cm  
  \begin{tabular*}{\hsize}{@{\hspace{0.15cm}}@{\extracolsep{\fill}}lcccccc@{\hspace{0.3cm}}}
    \hline
    \textbf{\(N\)} & \textbf{WB-PESQ} & \textbf{STOI} & \textbf{CSIG} & \textbf{CBAK} & \textbf{COVL} & \textbf{\# Params} \\
    \hline
    1  & 3.46 & \textbf{0.96} & 4.71 & 3.92 & 4.20  & 1.11M                       \\
    2  & 3.51 & \textbf{0.96} & 4.76 & 3.96 & 4.25  & 2.00M                       \\
    3  & 3.56 & \textbf{0.96} & \textbf{4.79} & 3.99 & \textbf{4.30}  & 2.89M        \\
    4  & \textbf{3.57} & \textbf{0.96} & \textbf{4.79} & \textbf{4.00} & \textbf{4.30}  & 3.78M   \\
    \hline
  \end{tabular*}
  \vspace{-0.3cm}
\end{table}

\subsection{Ablation Study}

The experiments in Table \ref{tab4} investigate the impact of varying the number of TS-Mamba blocks \(N\) within Mamba-SEUNet (M) on SE performance. As \(N\) increases from 1 to 3, there is a consistent improvement in objective performance metrics, including PESQ, STOI, and three MOS scores, demonstrating the efficacy of adding TS-Mamba blocks in enhancing speech quality. However, with a further increase in \(N\) to 4, while some metrics such as PESQ and CBAK exhibit slight improvements, others remain unchanged, indicating a potential plateau in performance. Moreover, as \(N\) increases, the number of model parameters also rises, highlighting the trade-off between performance gains and model complexity.

\begin{table}[h]
  \vspace{-0.3cm}
  \caption{Performance Comparison of Mamba, Transformer, and Conformer on the Same Architecture}
  \label{tab5}
  \centering
  \vspace{-0.1cm}
  \tabcolsep=0.11cm  
  \begin{tabular*}{\hsize}{@{\hspace{0.15cm}}@{\extracolsep{\fill}}lcccccc@{\hspace{0.3cm}}}
    \hline
    \textbf{Model} & \textbf{WB-PESQ} & \textbf{CSIG} & \textbf{CBAK} & \textbf{COVL} & \textbf{\# Params} & \textbf{FLOPs} \\
    \hline
    Mamba  & \textbf{3.57} & \textbf{4.79} & \textbf{4.00} & \textbf{4.30} & 3.78M & 10.28G  \\
    Conformer  & 3.45  & 4.70 & 3.91 & 4.19  & 2.94M  & 34.07G                  \\
    Transformer   & 3.52  & 4.76 & 3.96 & 4.25  & 4.55M & 51.69G                   \\
    \hline
  \end{tabular*}
  \vspace{-0.3cm}
\end{table}

\subsection{Mamba/Transformer/Conformer Block}

To ensure a fair comparison of Mamba with conformer and transformer, we employ the proposed U-Net architecture, replacing TS-Mamba with TS-Conformer as introduced in \cite{CMGAN} and TS-Transformer as described in \cite{TSTNN}, as shown in Table \ref{tab5}. The results indicate that Mamba outperforms conformer, achieving better performance with a slight increase in model parameters and a significant reduction in FLOPs, with a 0.12 improvement in the PESQ score. Compared to the transformer, which replaces the first linear layer in the feedforward network \cite{transformer} with gated recurrent units (GRU) \cite{GRU} to learn positional information, Mamba exhibits better performance, with a 0.05 improvement in the PESQ score, while requiring fewer parameters and lower FLOPs. These findings underscore the effectiveness of Mamba in SE tasks.

\section{Conclusions}
\label{Conclusions}

In this study, we introduce Mamba-SEUNet, a U-Net style SE network based on TS-Mamba blocks. This architecture leverages bidirectional Mamba blocks to effectively capture both past and future information, addressing the limitations of existing transformer-based methods. By integrating TS-Mamba blocks into the U-Net framework, Mamba-SEUNet enhances multi-scale information capture while maintaining computational efficiency. Experimental results on the VCTK+DEMAND dataset demonstrate that Mamba-SEUNet achieves state-of-the-art (SOTA) SE performance with low computational complexity. In the future, we aim to extend this network to other SE tasks and further optimize its performance on more challenging datasets.

\bibliographystyle{IEEEbib}
\bibliography{strings,refs}

\end{document}